# Generalized Dynamic Junction Theory to Resolve the Mechanism of Direct Current Generation in Liquid-Solid Interfaces

Cristal Solares-Bockmon[a], Aniqa Ibnat Lim[b], Mohammadjavad Mohebinia[a], Xinxin Xing[b], Tian Tong[b], Xingpeng Li[b], Steven Baldelli[c], T. R. Lee[c,d], Wei Wang[e], Zhaoping Liu[e], Jiming Bao[*, a, b, c, d, f]

[a]Materials Science and Engineering, University of Houston, Houston, Texas 77204, USA

[b]Department of Electrical and Computer Engineering, University of Houston, Houston, Texas 77204, USA

[c]Department of Chemistry, University of Houston, Houston, Texas 77204, USA

[d]Texas Center for Superconductivity at the University of Houston, University of Houston, Houston, Texas 77204, USA

[e]Key Laboratory of Graphene Technologies and Applications of Zhejiang Province, Ningbo Institute of Materials Technology and Engineering, Chinese Academy of Sciences, Ningbo, Zhejiang 315201, China

[f]Department of Physics, University of Houston, Houston, Texas 77204, USA

**Corresponding Authors**

*E-mail: jbao@uh.edu.

# Abstract

Despite the unsettled mechanism of electricity generation from the continuous flow of liquids on a surface, the charge-discharge theory has been widely accepted for alternating current (AC) generation from a moving droplet. It has been recently extended to rationalize direct current (DC) generation across a droplet moving between two different materials. By designing a reconfigurable contact between a metal wire and a water droplet moving on graphene, we show that the charge-discharge theory cannot explain the reversal of current when water-metal interfaces switch from dynamic to static. All experiments can be described after we distinguish a dynamic from a static interface and generalize the photovoltaic-like effect to all dynamic junctions: excited electrons and holes in a moving interface will be separated and swept under the built-in electrical field, leading to a DC response. This generalized theory will lead to an understanding and the design of efficient electricity generation based on interfacial charge transfer.

1. Introduction

Harvesting electricity from fluid flow on a solid surface through interfacial charge transfer has attracted enormous attention because of the potential utilization of ubiquitous water and a new generation of hydroelectricity. However, the underlying mechanisms remain elusive and under intensive debate while key experiments are still difficult to reproduce and quantify [1-3]. For example, a DC response from flowing water on graphene was reported in 2011 [4]. Still, this observation was quickly questioned and was believed to come from metal contacts because no current was observed when metal contacts to graphene were well isolated [5]. In the meantime, many theories have been proposed such as momentum transfer, coulomb dragging and streaming potential [1, 3, 6-10]. In contrast, the electricity generation from a moving droplet on graphene was observed and rationalized by a charge-discharge theory: electrons will be attracted to the front of the droplet and released from the rear of the droplet [11]. In other words, the free charges will be shuffled from one side of the material surface to the other side by the moving droplet, leading to a current. The observation was quickly reproduced in many interfaces between different liquids and various 2D materials [2, 9, 12-21]. Although the charge-discharge theory is fundamentally different from the theories for electricity generation from the flowing liquids [2, 3, 9], the community has accepted it to describe all droplet-induced electricity [2, 9, 10, 12-21].

Because the current direction depends on the moving direction of a droplet, an alternating current (AC) is generated when the droplet moves back and forth. The generation of a direct current (DC) is desirable for low-power portable electronic devices; thus, it has attracted enormous interest and has been reported for various liquid-solid interfaces [22, 23]. The charge-discharge theory is a natural choice to describe the DC between two materials sandwiching a moving droplet. It states that the DC is due to the recombination of opposite charges released from two materials with different Fermi levels [23, 24]. However, a competing theory is based on dynamic water-silicon, which states that excited electrons and holes in the depletion region of silicon will be separated by a built-in electric field and become a DC [22, 25]. This mechanism is similar to photovoltaics (PV) except that electrons and holes are excited by interfacial mechanical movement instead of incident photons. While these two competing views are fundamentally different, they seem to work well because they deal with various experiments, one with a single dynamic liquid-solid interface, while the other deals with two dynamic liquid-solid interfaces.

The resolution of the controversial mechanisms will lead to a better understanding of dynamic liquid-solid interfaces and facilitate the development of more efficient power generators. In this work, we design new generators with a unique liquid-solid interface that can be configured as a dynamic or static junction. In this way, both theories can be applied to the devices and systematically tested. After demonstrating the failure of the charge-discharge theory, we will generalize the PV-like DC generation theory and use it to rationalize our and other reported experiments.

## 2. Results and Discussion

Figs. 1a-b show a DC generator schematic and picture with a water droplet between graphene and a suspended metal wire. We used a mica plate and a motor to drag the droplet freely on graphene. We chose Au and Al because they possess work functions below and above that of graphene. Large-area single-layer graphene was synthesized by chemical vapor deposition and transferred to an A-4 size PET film so that a single-layer graphene strip can be cut into any length and width.

This device is similar to DC generators described by the charge-discharge theory [23, 24]. As the droplet moves, it forms two dynamic junctions: water-graphene and water-metal. The current is measured with a Keithley 2400 source meter. Figs. 1c-d show that the generated current goes from Au to graphene when an Au wire is used, but the current goes from graphene to Al with an Al wire. This material-dependent DC is due to the difference in the work function of Au and Al. The current's direction can be described by charge-discharge theory: the current flows from a higher work function material to a lower work function electrode [23, 24].

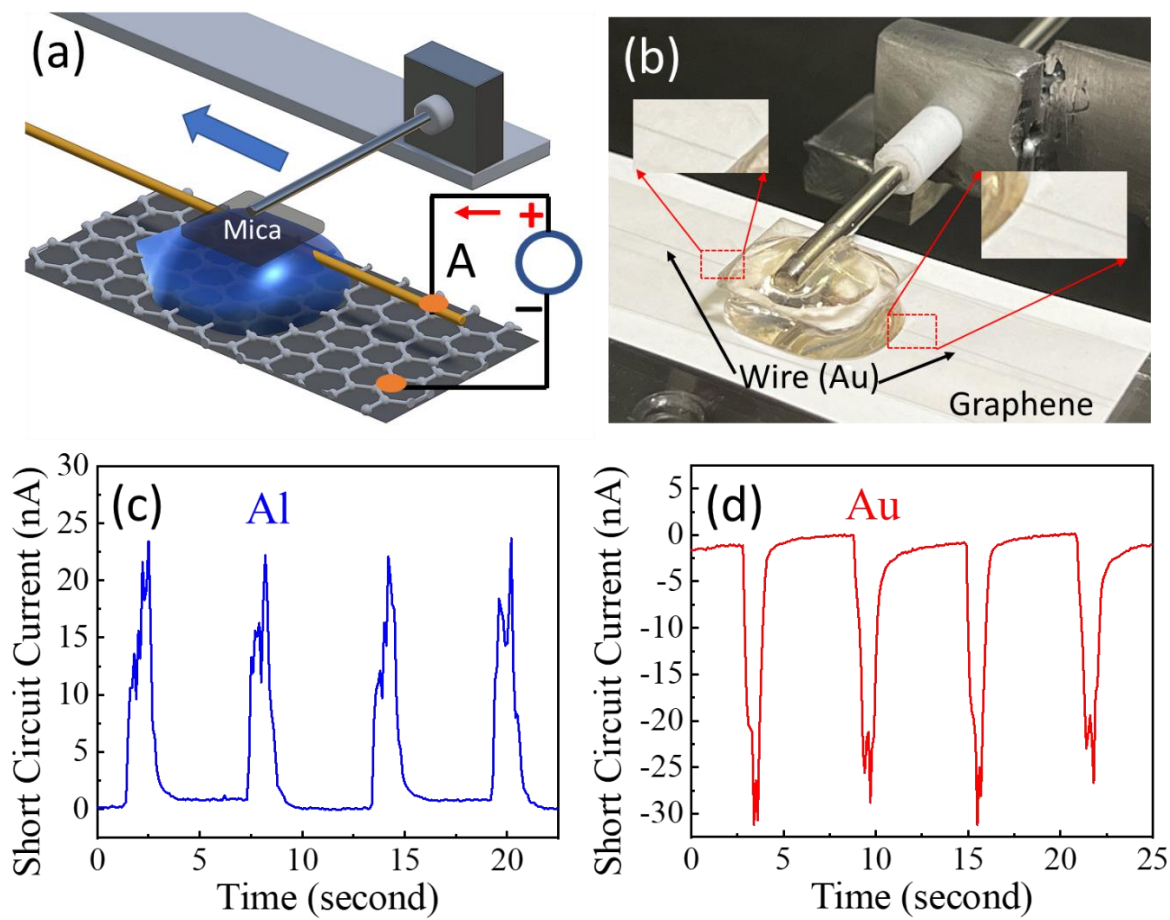

**Fig. 1.** DC between graphene and a suspended metal wire with a moving water droplet (a) Schematic of the experimental setup. (b) Picture of the setup with an Au wire. (c-d) The current between graphene and (c) Al wire and (d) Au wire when the water droplet moves back and forth. A positive current indicates current from graphene to wire through the current meter.

We then inserted the same wire in a capillary tube and let the tube-wire move along with the droplet. Fig. 2a shows the schematic, and Fig. 2b shows a picture of the setup. A DC is also observed when the droplet moves, but Figs. 2c-d show that the DC is significantly reduced with the Al wire compared with that of the suspended Al wire, and the current changes direction for the Au wire. Note that in this case, the water is not moving on the surface of the wire since both the wire and water are tightly confined in the capillary tube, so the water-metal interface is static. In other words, the DC generator has only one dynamic junction between the water and graphene instead of two junctions with a suspended wire, so the change in current direction is unsurprising.

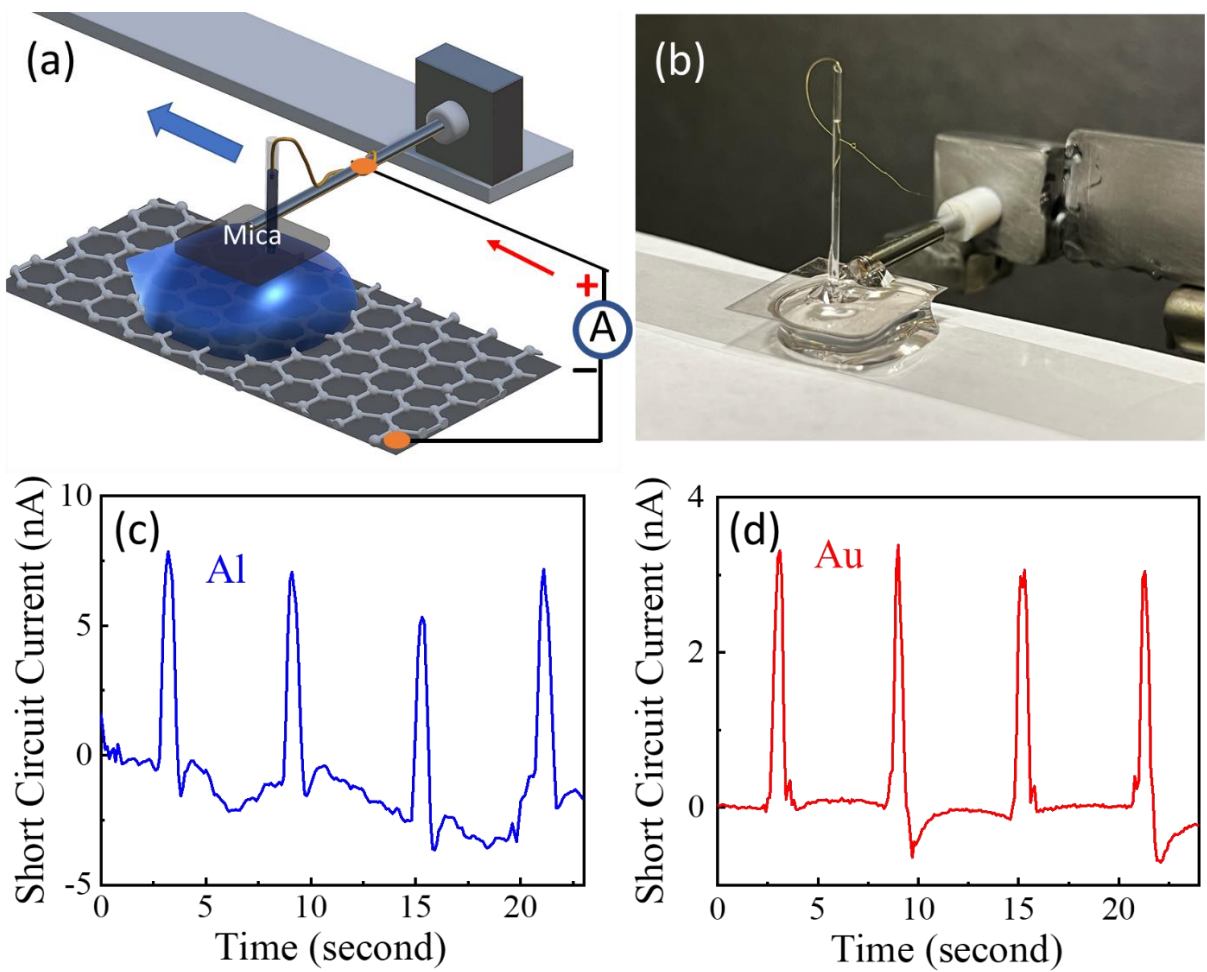

**Fig. 2.** DC between graphene and a metal wire moving with a droplet. (a) Schematic of the experimental setup. (b) Picture of the setup with an Au wire. (c-d) The current between graphene and (c) Al wire and (d) Au wire when the water droplet moves back and forth.

We argue that no matter whether the devices in Fig. 2 are treated as single or double dynamic junctions, the charge-discharge theory cannot explain the observed currents. According to the charge-discharge theory [23, 24], their work functions will determine the induced free charges and the DC current between graphene and wires.

Fig. 3a shows the work functions (Fermi levels) of graphene, water, Au, and Al. The Fermi levels of graphene, Au, and Al are based on Kelvin probe measurments, while the Fermi level of water (4.6 eV) is based on literature [22, 26, 27]. Because Al possesses a lower work function than graphene, Fig. 3b shows that free electrons will be induced on the Al wire, and an equal amount of free holes will be induced on graphene. As the droplet moves, holes will be released from graphene to the external circuit; free electrons will be ejected from Al; and electrons and holes then recombine through the external circuit and are detected as a positive DC. When Au is used, free holes instead of electrons will be induced on Au because of its higher work function. Consequently, the free charges on graphene must change and become negative, as does the DC, as shown in Fig. 3c and Fig. 1.

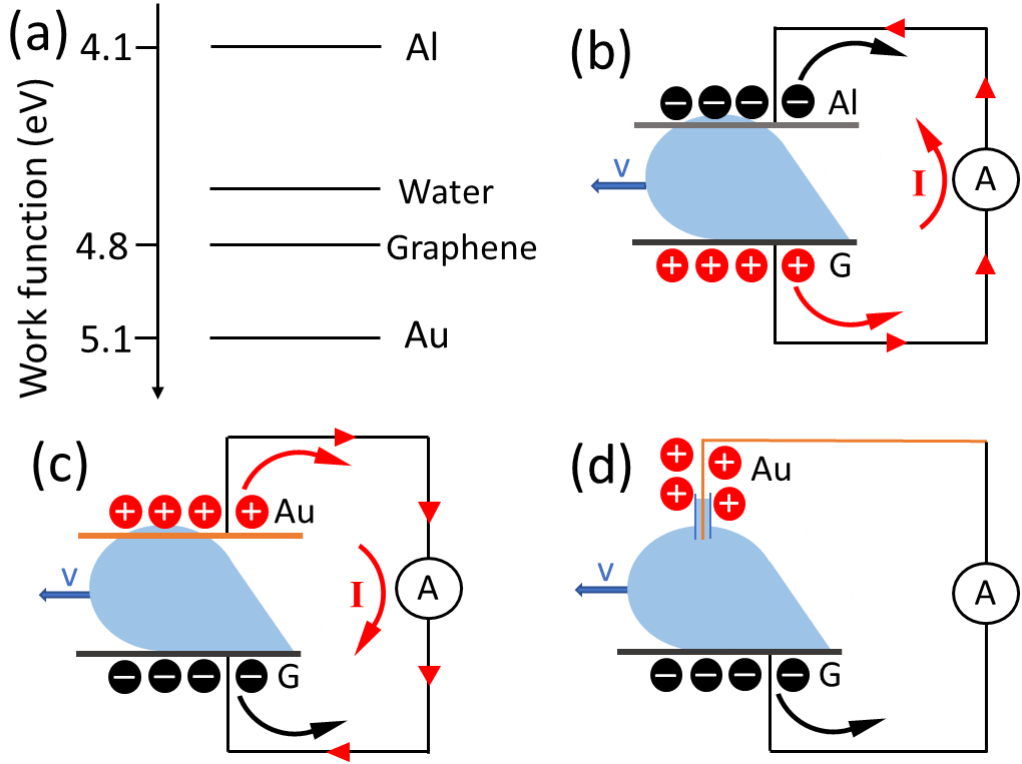

**Fig. 3.** Failure of charge-discharge theory. (a) Fermi levels of graphene, water, Al, and Au. (b-c) Successful charge-discharge description of DC between graphene and suspended (b) Al wires and (c) Au wires. (d) Expected DC from the charge-discharge theory when applied to the moving Au wire.

Following the same logic, we show the same induced free charges in Fig. 3d when the Au wire is placed in the capillary tube. Since the water surrounding the wires remains stationary even when the droplet moves, the water-Au interface remains a static junction, so holes will not be released or discharged from the Au wire. On the contrary, the free electrons on graphene should be discharged when the droplet moves. Because there is no equal amount of free holes from the Au wire to recombine with the free electrons from graphene, the current should either be zero or reduced to half if electrons alone are released to the external circuit and contribute to the DC. In either case, our observations in Fig. 2d contradict the prediction from the charge-discharge theory. For Al, although the current flows in the same direction, it is reduced to one third, lower than the expected half when holes from graphene contribute only to the current.

An alternative theory for the DC generation is based on a moving liquid-solid interface. The DC from a moving water droplet on a silicon substrate was reported and rationalized by a photovoltaic-like theory: electrons and holes are excited by the moving interface in the depletion region of Si and then separated by the built-in field, leading to a short-circuit current [22, 25]. Besides liquid-solid interfaces, DC generations have been observed in many solid-solid interfaces or junctions, including traditional Si [28-33], III–V compounds [32], carbon materials [34], and new materials such as perovskites and 2D nanomaterials [35]. Although a detailed microscopic theory is still lacking, especially the mechanism of carrier excitation, the direction of DC is universal and can be determined by the built-in electric field in the junctions. Here we propose that DC can be generated in all dynamic junctions and determined by the built-in field. This generalized dynamic junction theory is valid for our liquid-metal and liquid-graphene interfaces. In the following paragraph, we provide a brief description of the principle of the theory and use it to rationalize our experimental observations.

Figs. 4a-b show that when two materials with different Fermi levels are brought into contact, interfacial charge transfer occurs to align their Fermi levels. This charge transfer will create a space charge region with a built-in field and voltage to prevent further net charge transfer. As a static junction in thermal equilibrium, it will not produce any current or voltage for the external circuit. This is a general principle of physics regardless of materials. Such a junction serves as an electrical contact between two materials in most cases. However, in the case of a relative mechanical movement between two materials as shown in Fig. 4c, the junction becomes dynamic so that extra electrons and holes will be generated and separated by the built-in field, producing current to the external circuit. Figs. 4d-f show the static junctions of water with Al, Au, and graphene based on their Fermi levels. Note that these junctions are local between two materials only; thus, the notion that the induced free charge on graphene in water is switched between Al and Au wires in Figs. 3c and 3d is incorrect from the very beginning.

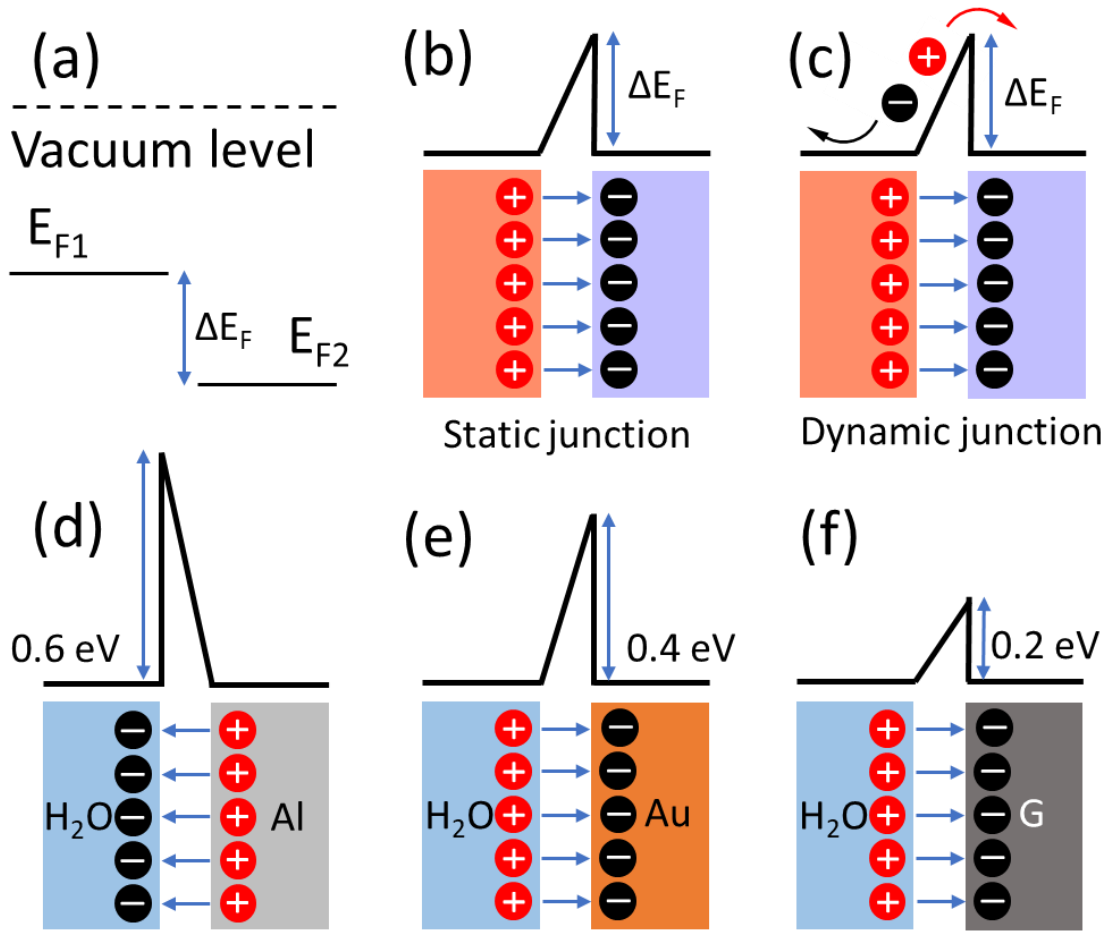

**Fig. 4.** Generalized dynamic junction theory for DC generation. (a) Two materials with different Fermi levels. (b) Space charge, built-in electric field and potential after two materials are in contact. (c) Separation of electron and hole in a dynamic junction to generate a DC. (d-f) Schematics of space charge and built-in potentials at the interfaces of (d) water-Al, (e) water-Au, (f) water-graphene based on their Fermi levels in Fig. 3a. G indicates graphene.

Having established junctions among water, graphene, Au/Al and understanding the difference between a static and a dynamic junction, we can use them to construct multi-junction generators and understand their fundamental operations. Figs. 5a-b show the built-in fields and energy diagrams for graphene-water-Al DC generators in Figs. 1c and 2c. Because of their relative Fermi levels, two built-in fields are aligned in the same direction. Both junctions are dynamic for the suspended Al wire in Fig. 1c and contribute to the overall current. While for the moving Al wire in Fig. 2c, the water-Al junction is static as an electrical contact, and only the graphene-water junction contributes to the current; these factors explain why both currents flow in the same direction, but the current with the suspended Al wire is larger than the current with the moving Al wire. The case for the Au wires is different: two built-in fields are in the opposite directions, as shown in Figs. 5c-d. Because the built-in voltage in the water-Au junction is much larger than that in the water-graphene junction, the current from the water-Au junction dominates, giving rise to a net DC from Au to graphene in the case of the suspended Au wire in Fig. 1d. But for the moving Au nanowire in Fig. 2d, the current is determined by the weaker water-graphene junction only; the current goes in the opposite direction despite being much smaller.

The failure of the charge-discharge theory for the DC generators can also help us understand its failure for the moving droplet AC generators [2, 9, 11-21]. A major assumption is that there is no charge transfer between the droplet and its contacting solid surface, and the role of the droplet is to shuffle the free charges from the front to the rear side of the droplet on the surface. This assumption directly contradicts our observations. There is a DC between water and graphene, and we believe that such current is general and will exist in any liquid-solid dynamic junctions. On the other hand, since the original model is highly phenomenological, it still cannot provide a clear microscopic picture and quantitative results although it has been refined many times over the past years [11, 15, 18, 19].

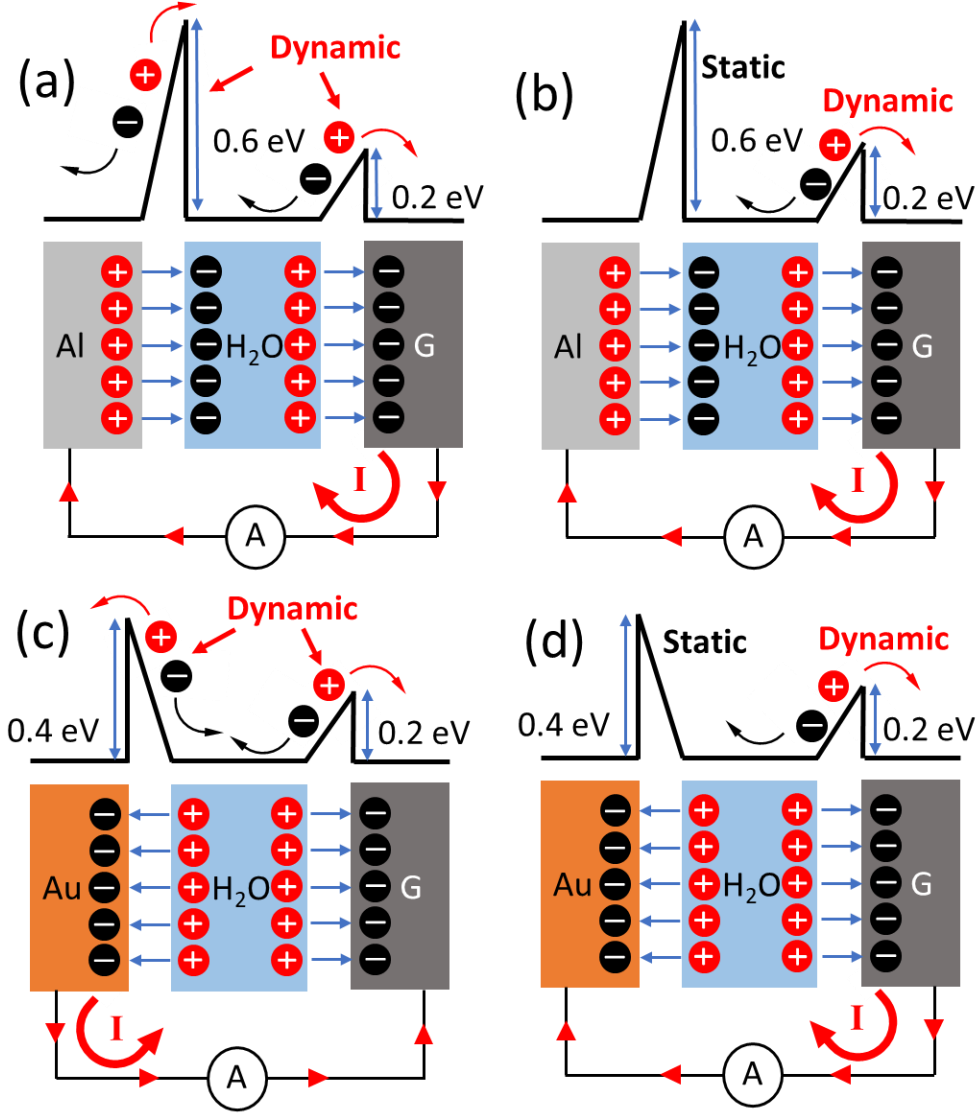

**Fig. 5.** Application of generalized dynamic junction theory to DC generators in Figs. 1 and 2. (a-b) Current generations in graphene-Al generators with (a) suspended and (b) moving Al wires. (c-d) Current generations in graphene-Au generators with (c) suspended and (d) moving Au wires.

## 3. Conclusions

In summary, we designed unique droplet DC generators with switchable static/dynamic junctions; we then used them to show the failure of the charge-discharge theory. We proposed that a DC will be generated in any dynamic junction and then provided supporting experimental evidence. We also pointed out that the well-accepted charge-discharge theory is insufficient to explain the droplet

AC generators, and that our generalized dynamic junction theory can be applied to other junctions and interfaces.

## Acknowledgments

Funding: J.M.B. and T.RL acknowledge support from the Robert A. Welch Foundation (E-1728 J.M.B. and E-1320 T.R.L.) and the Texas Center for Superconductivity at the University of Houston. T.R.L also acknowledges support from the National Science Foundation (CHE-2109174). Z.P.L. acknowledges support from  Innovation 2025 Major Special Program of Ningbo (Grant No. 2019B10097).